\documentclass[12pt]{nature}

\usepackage{txfonts}
\usepackage{graphicx}
\usepackage{lineno}
\usepackage{caption}
\usepackage{fancyhdr}

\title{Universal dynamo scaling in young and evolved stars}

\author{Jyri J. Lehtinen$^{1,2}$, Federico Spada$^1$, Maarit J. K\"apyl\"a$^{2,1}$, Nigul Olspert$^1$ \& Petri J. K\"apyl\"a$^{3,2}$}

\begin{document}

\thispagestyle{empty}

\noindent
This is the initial submission version of the paper \\
\textit{Common dynamo scaling in slowly rotating young and evolved stars} (Lehtinen et al.~2020). \\

\noindent
The final accepted version can be found at \\
\texttt{https://www.nature.com/articles/s41550-020-1039-x}

\pagebreak

\setcounter{page}{1}

\maketitle

\begin{affiliations}
\item Max Planck Institute for Solar System Research, 37077 G\"ottingen, Germany
\item Department of Computer Science, Aalto University, PO Box 15400, FI-00076 Aalto, Finland
\item Institut f\"ur Astrophysik, Georg-August-Universit\"at G\"ottingen, 37077 G\"ottingen, Germany
\end{affiliations}

\begin{abstract}
Activity caused by surface magnetism is a pervasive feature of cool late-type stars where a dynamo mechanism is supported in the outer convective envelopes of the stellar interiors. The detailed mechanism responsible for this dynamo is still debated but its basic ingredients include convective turbulence and non-uniformities in the stellar rotation profile\cite{Charbonneau2010Dynamo}, although the role of the former divides opinions. Both the observed surface magnetic fields\cite{Reiners2009MagneticRotationActivity} and activity indicators from the chromosphere\cite{Noyes_ea1984} to the transition region\cite{Gunn1998RotationActivity} and corona\cite{Wright2011RotationActivity} are known to be closely connected with the stellar rotation rate. These relationships have been intensely studied in recent years, some works indicating that the activity should be related to the Rossby number\cite{Wright2011RotationActivity,Mamajek2008RotationActivity} , quantifying the stellar rotation in relation to the convective turnover time, while others claim that the tightest correlation can be achieved using the rotation period alone\cite{Pizzolato2003Activity,Reiners2014Activity}. Here we tackle this question by including evolved giant stars to the analysis of the rotation-- activity relation. These stars rotate very slowly compared to the main sequence stars, but still show strikingly similar activity levels\cite{Strassmeier1990GiantRotationActivity}. We show that by using the Rossby number, the two stellar populations fall together in the rotation--activity diagram, and follow the same activity scaling. This suggests that turbulence has a key role in driving stellar dynamos, and that there appears to be a universal turbulence-related dynamo mechanism explaining magnetic activity levels of both main sequence and evolved stars.
\end{abstract}

A common practice in studies of stellar activity\cite{Noyes_ea1984,Gunn1998RotationActivity,Wright2011RotationActivity,Mamajek2008RotationActivity} and magnetism\cite{Reiners2009MagneticRotationActivity,Vidotto2014MagneticRotationActivity} has been to assume that the scaling of the activity level is best explained by the Rossby number ${\rm Ro} = P_{\rm rot}/\tau_{\rm c}$, which is the ratio between the stellar rotation period $P_{\rm rot}$ and the convective turnover time $\tau_{\rm c}$. This interpretation has, however, been contested recently\cite{Pizzolato2003Activity,Reiners2014Activity}. The main point of criticism against the use of the Rossby number has been that determining its value requires the knowledge of the convective turnover time in the interior of a star. Since this is not a directly observable quantity, it has to be estimated instead using stellar structure models\cite{Gilliland:1985} or empirical fits\cite{Noyes_ea1984}, which increases the risk of introducing systematic errors in the analysis. On the main sequence stars the coronal X-ray luminosity has been noted to be empirically correlated with the rotation period\cite{Pizzolato2003Activity} as $L_{\rm X} \propto P_{\rm rot}^{-2}$. Equivalently, the ratio of the X-ray to bolometric luminosity has been related to the rotation period and stellar radius as $L_{\rm X}/L_{\rm bol} \propto P_{\rm rot}^{-2}R^{-4}$, resulting in a marginally better fit than when relating $L_{\rm X}/L_{\rm bol}$ to empirical $\rm Ro$ of the same stars\cite{Reiners2014Activity}.

Resolving this controversy on the capability of the Rossby number for correctly describing the activity scaling is important, as it gives direct clues on the dynamo type that is operational in the stellar convection zones. Stellar dynamos depending solely on the rotation period could indicate preference for Babcock-Leighton-type dynamos, where a crucial part of the dynamo action relies on rising flux tubes becoming twisted by the Coriolis force due to rotation\cite{Babcock61,Leighton69,Dikpati1999FluxTransport}. The rival theory of turbulent dynamos relies on the generation of magnetic fields by rotationally affected convective cells\cite{Parker1955}. In this case the activity level should also critically depend on the properties of the turbulence, such as the convective turnover time. Hence, the relevant parameter for such dynamos is the Rossby number which describes the rotational influence on convective turbulence\cite{Brandenburg2005NonlinearDynamo}.

Stars develop increasingly thick convective envelopes as they evolve off the main sequence. As a result the value of $\tau_{\rm c}$ will start to increase once the star reaches the end of its main sequence phase and starts to turn into a giant\cite{Gilliland:1985}. As $P_{\rm rot}$ will continue to increase over time\cite{Skumanich1972Rotation} due to magnetic braking and expanding stellar radius, the values of $\rm Ro$ and $P_{\rm rot}$ evolve differently after the end of the main sequence phase. This offers a possibility to test which of the proposed rotation--activity scaling relations works the best once both main sequence and evolved stars are studied in conjunction.

\begin{figure}
\centering
\includegraphics[width=\linewidth]{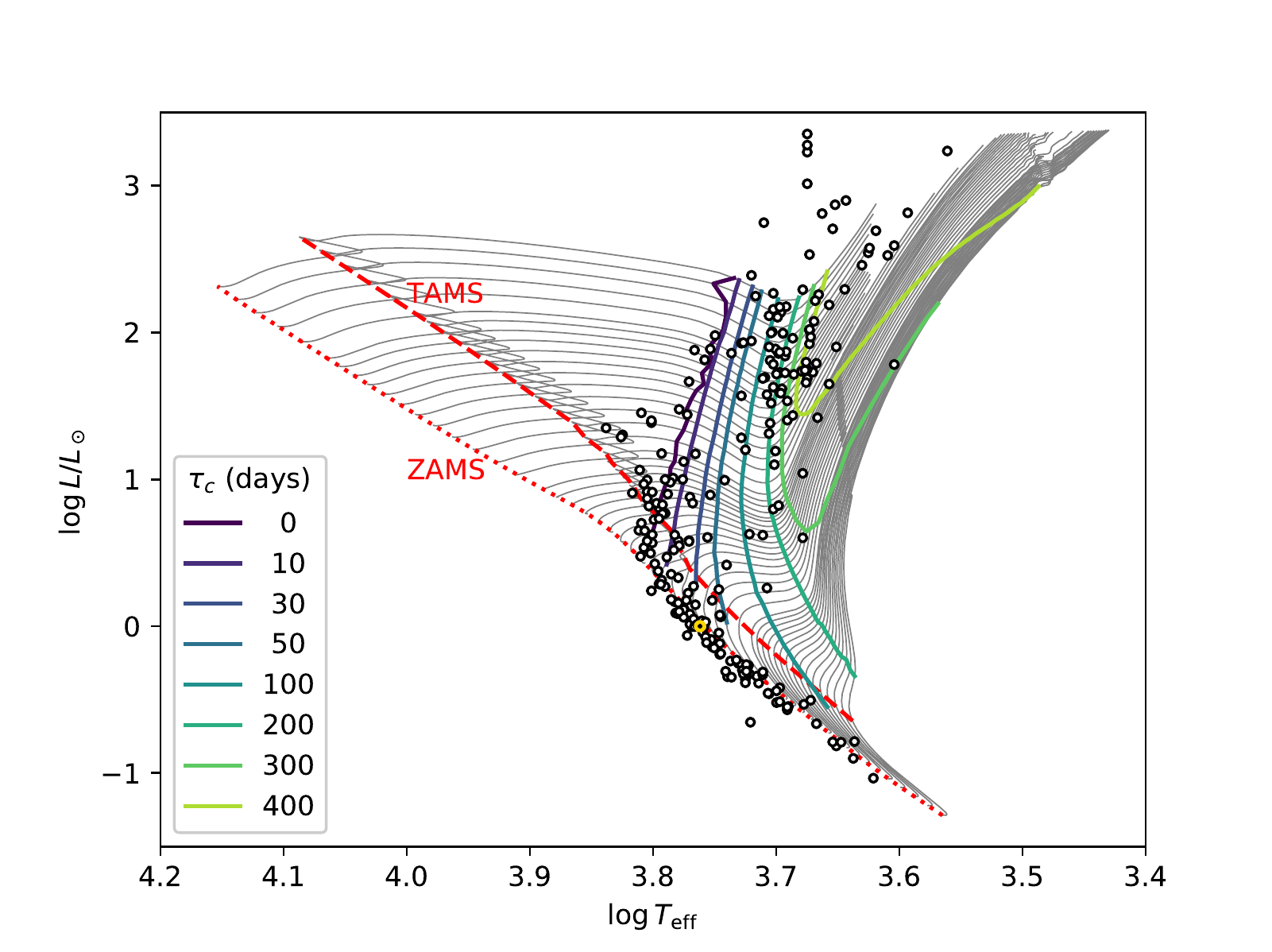}
\caption{\textbf{Hertzsprung-Russell diagram of the stellar sample.} The stars are shown together with a subset of the model evolutionary tracks used to determine their convective turnover times, $\tau_{\rm c}$. Only the solar metallicity tracks are shown here for clarity. The red dotted and dashed lines mark the zero age main sequence (ZAMS) and the termination age main sequence (TAMS), respectively, while $\tau_{\rm c}$ isocontours are shown as solid colored lines. The Sun is indicated by the yellow circle.}
\label{hr-fig}
\end{figure}

A suitable data set for this study is provided by the chromospheric time series collected during the Mount Wilson Observatory (MWO) Calcium HK Project\cite{Wilson1978MWOHK}. These data allow straightforward derivation of $P_{\rm rot}$ and the average ratio of the Ca II H\&K line core emission to bolometric flux, $R'_{\rm HK} = F'_{\rm HK}/F_{\rm bol}$. This activity index, commonly expressed in the logarithmic form $\log R'_{\rm HK}$, quantifies the efficiency with which the full energy output of a star is converted to chromospheric heating. The turnover times, $\tau_{\rm c}$, needed for calculating $\rm Ro$, were derived from stellar structure models\cite{Spada_ea2017}, by fitting the model evolutionary tracks to the observed parameters of each star. The Hertzsprung-Russell diagram of the stellar sample is shown in Figure \ref{hr-fig} together with the model tracks and isocontours of the resulting $\tau_{\rm c}$ values.

\begin{figure}
\centering
\includegraphics[width=.7\linewidth]{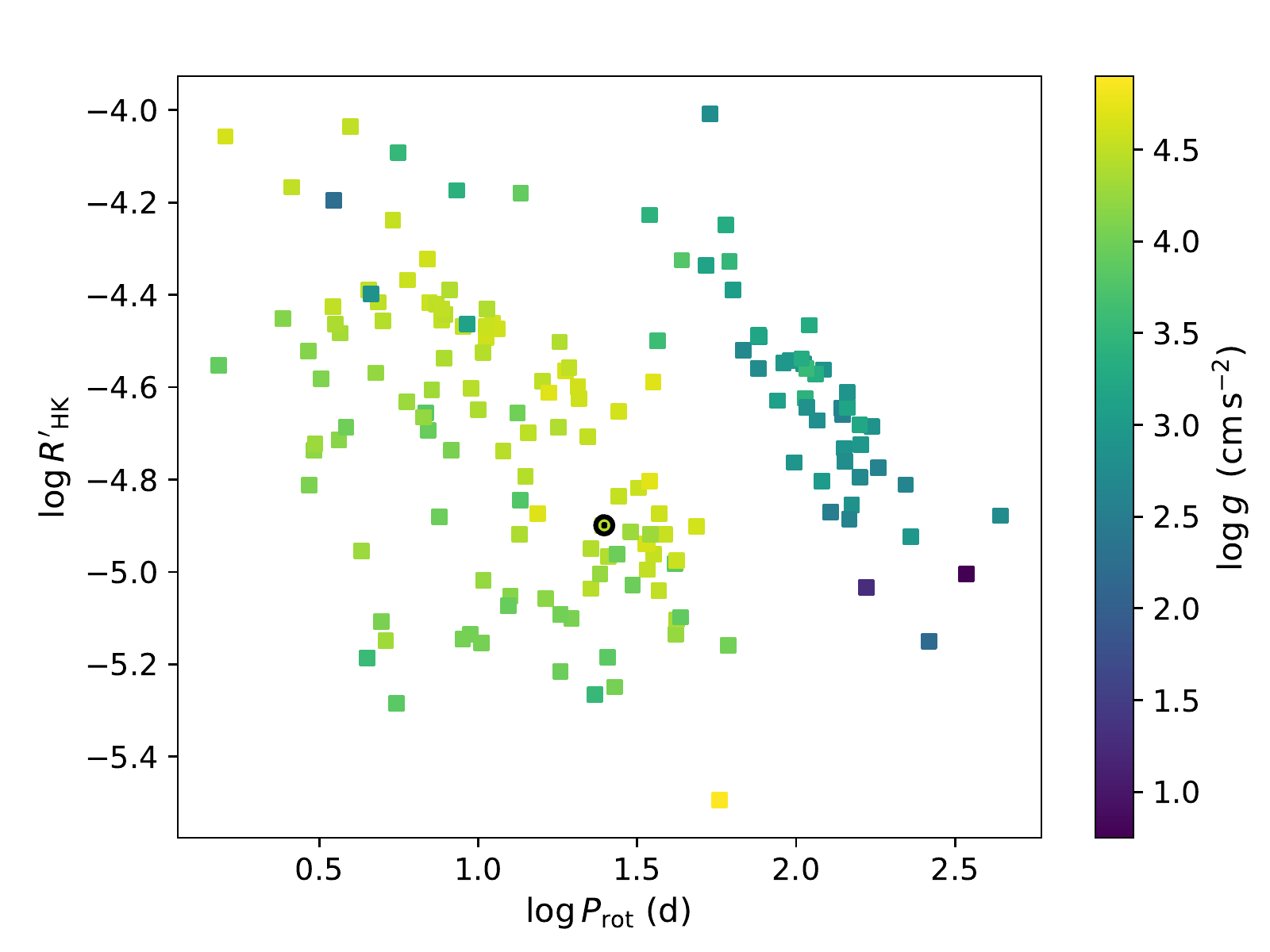}
\put(-355,235){\textbf{a}} \\
\includegraphics[width=.7\linewidth]{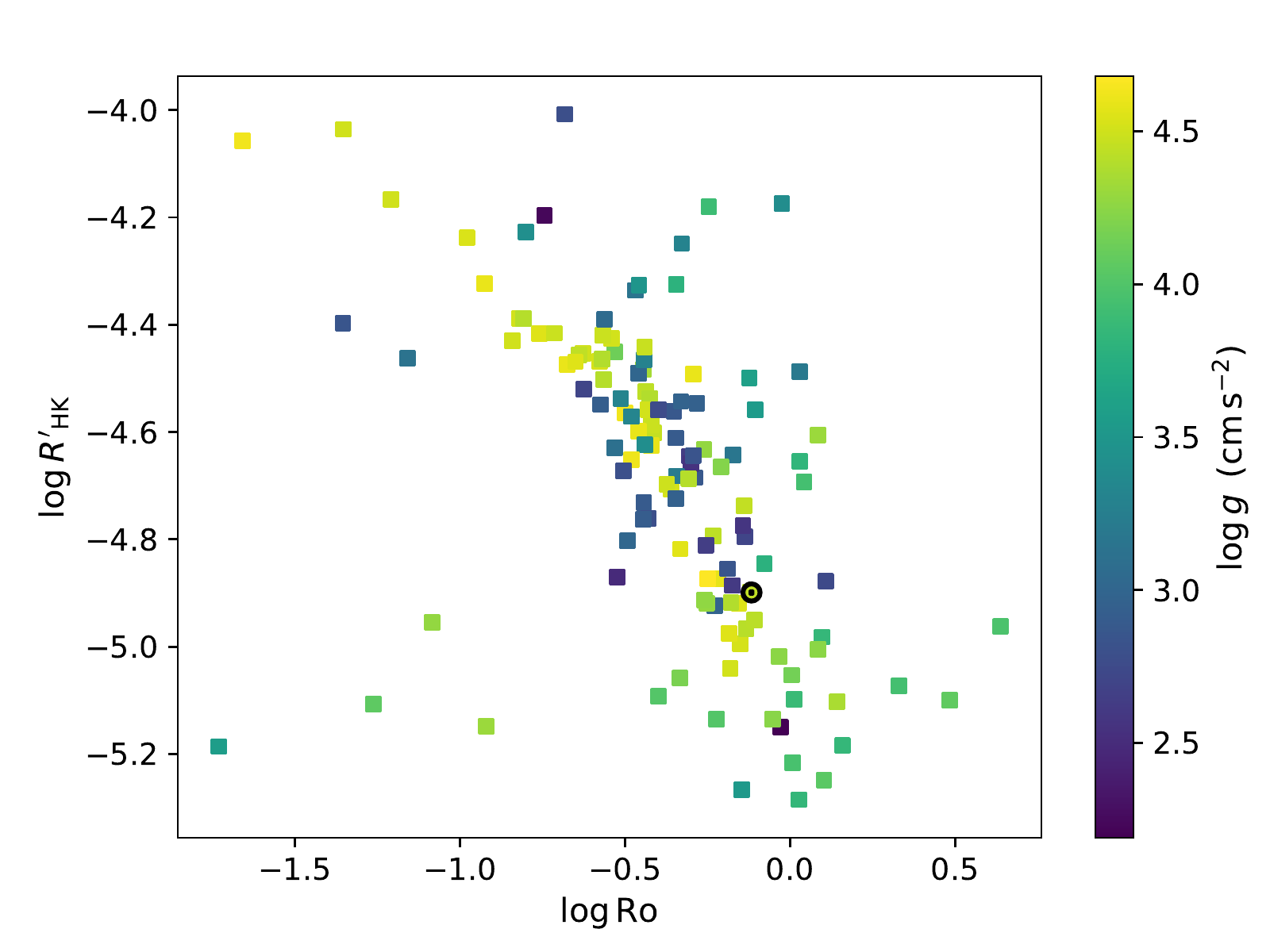}
\put(-355,235){\textbf{b}}
\caption{\textbf{Rotation--activity relation for main sequence and evolved stars.} \textbf{a}, Chromospheric activity, $\log R'_{\rm HK}$, vs.~rotation period, $P_{\rm rot}$. \textbf{b}, Chromospheric activity, $\log R'_{\rm HK}$, vs.~Rossby number, ${\rm Ro} = P_{\rm rot}/tau_{\rm c}$. The colour scale denotes the surface gravity, $\log g$, of the stars, distinguishing main sequence stars (yellow to light green) from the evolved stars (dark green to blue).}
\label{rotact-fig}
\end{figure}

In Figure \ref{rotact-fig} we show the emission ratio $\log R'_{\rm HK}$ of the MWO stars against both $P_{\rm rot}$ and $\rm Ro$. The stars are colour-coded by their observed surface gravity, $\log g$, to indicate their evolutionary phase, main sequence stars having high $\log g$ and the evolved stars low $\log g$. Against $P_{\rm rot}$ the stars are clearly split into two distinct groups, with the evolved stars showing periods of 10 to 100 times longer than main sequence stars of comparable activity levels. Likewise, within the main sequence there is substantial scatter between the least and most massive stars, as seen from the gradient between the least massive ones at $\log g > 4.5$ and the most massive ones at $\log g < 4.0$. When displayed against $\rm Ro$, on the other hand, the spread disappears and the stars collapse into a single rotation--activity sequence, irrespective of their mass or evolutionary stage.

\begin{figure}
\centering
\includegraphics[width=\linewidth]{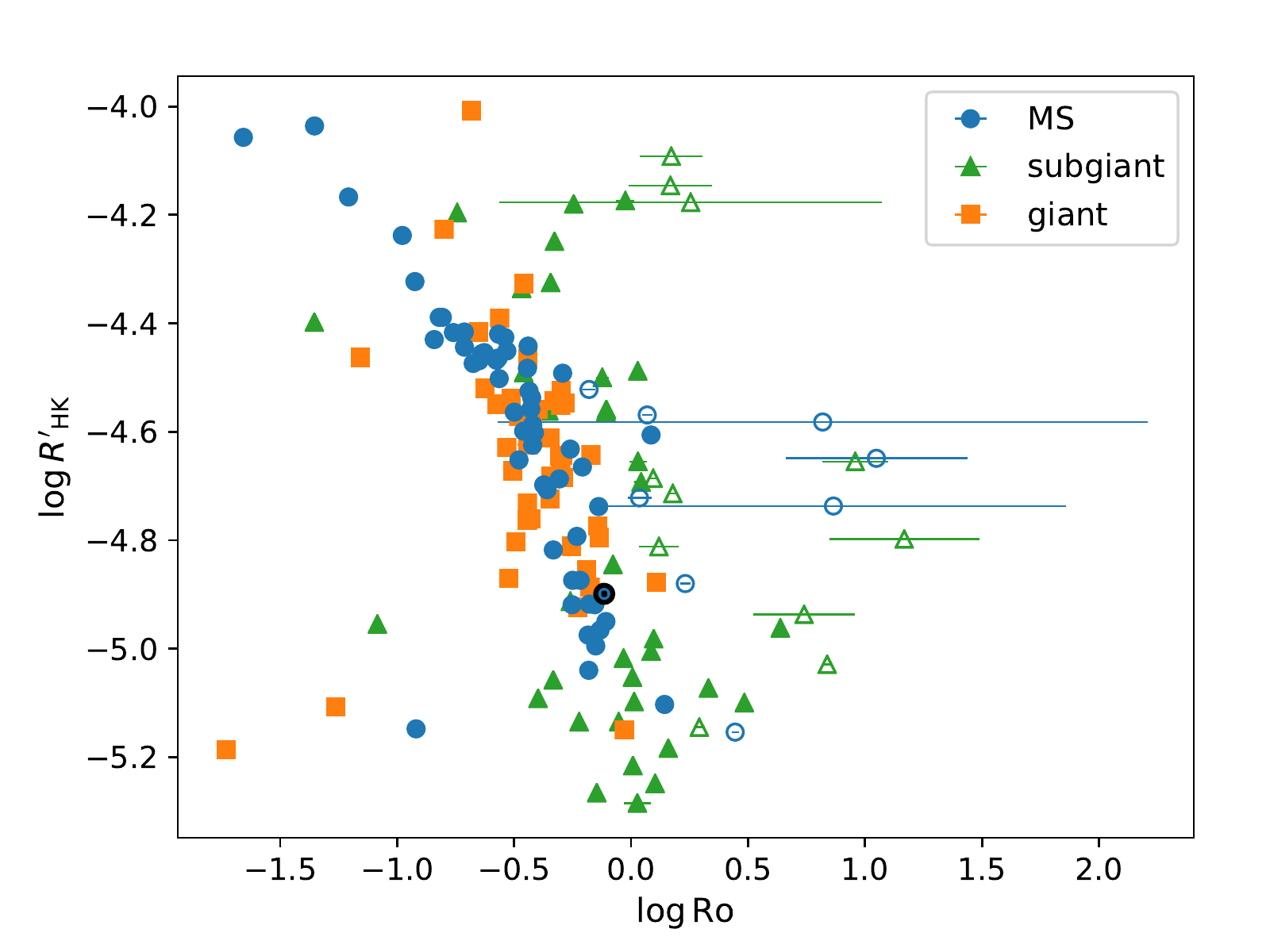}
\caption{\textbf{Rotation--activity relation including uncertainties for Rossby number.} Stars excluded from the analysis due to the uncertainty in recovering short turnover times ($\tau_{\rm c} < 5$ d) are shown with open symbols. The error bars indicate the estimated $\rm Ro$ uncertainties and different plot symbols are used for the main sequence (MS), subgiant, and giant stars.}
\label{evo-fig}
\end{figure}

There is still residual scatter remaining in the calculated $\rm Ro$ values, which needs to be discussed before evaluating the performance of the Rossby number as an activity scaling parameter. This scatter can be attributed largely to model uncertainties in deriving the $\tau_{\rm c}$ values from the structure model fits. In Figure \ref{evo-fig} we show $\log R'_{\rm HK}$ against $\log \rm Ro$ for all the stars for which an evolutionary track could be fitted. There are a considerable amount of outliers with sizable uncertainties to the large $\rm Ro$ side of the main bulk of the stars. These points mostly correspond to the stars with the shortest convective turnover times at $\tau_{\rm c} < 5$ d, as shown by the open symbols in Figure \ref{evo-fig}. The stars in question are located in the Hertzsprung-Russell diagram very near to the limit where the outer convection zone first appears as intermediate-mass stars ($\approx 2M_\odot$) evolve from the main sequence to the subgiant phase (cf.~the $\tau_{\rm c} = 0$ d contour in Figure \ref{hr-fig}). This limit remains poorly constrained, and thus the derived thickness of the convection zone near it is highly sensitive to model uncertainties and errors in the stellar parameters. For this reason we excluded the stars with $\tau_{\rm c} < 5$ d from our analysis as unreliable points. The scatter among the remaining stars is related largely to the subgiant phase, as can be expected due to the short evolutionary timescale of the depth of the convection zone and $\tau_{\rm c}$ during these phases\cite{Gilliland:1985}. The main sequence and giant stars are in much better agreement with each other.

The capability of the different parameters to explain the rotation–activity scaling can be further investigated through Gaussian clustering by finding the optimal configuration of bivariate Gaussian distributions to describe the data. Against $P_{\rm rot}$ this analysis leads to two distinctly separate clusters for the main sequence and evolved stars, while against $\rm Ro$ it favours a single narrow cluster with a broad overlapping secondary cluster, related to the more uncertain $\tau_{\rm c}$ values (Extended Data Figure \ref{cluster-fig} a,b). Most of both the main sequence and evolved stars fall neatly inside the narrow core cluster, which demonstrates that the Rossby number is indeed a sufficient parameter for explaining the activity scaling of both of these evolutionary stages.

\begin{figure}
\centering
\includegraphics[width=.7\linewidth]{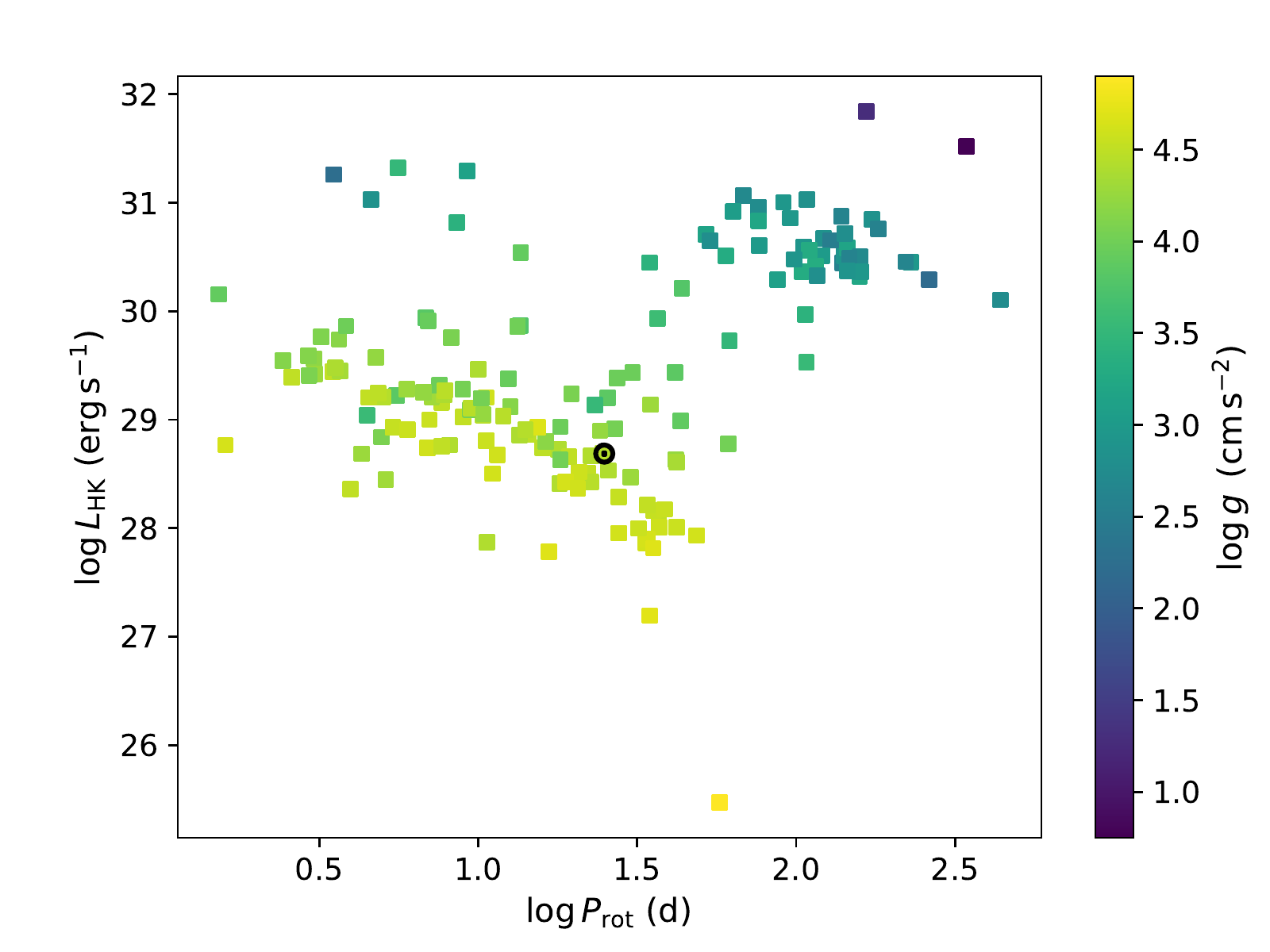}
\put(-355,235){\textbf{a}} \\
\includegraphics[width=.7\linewidth]{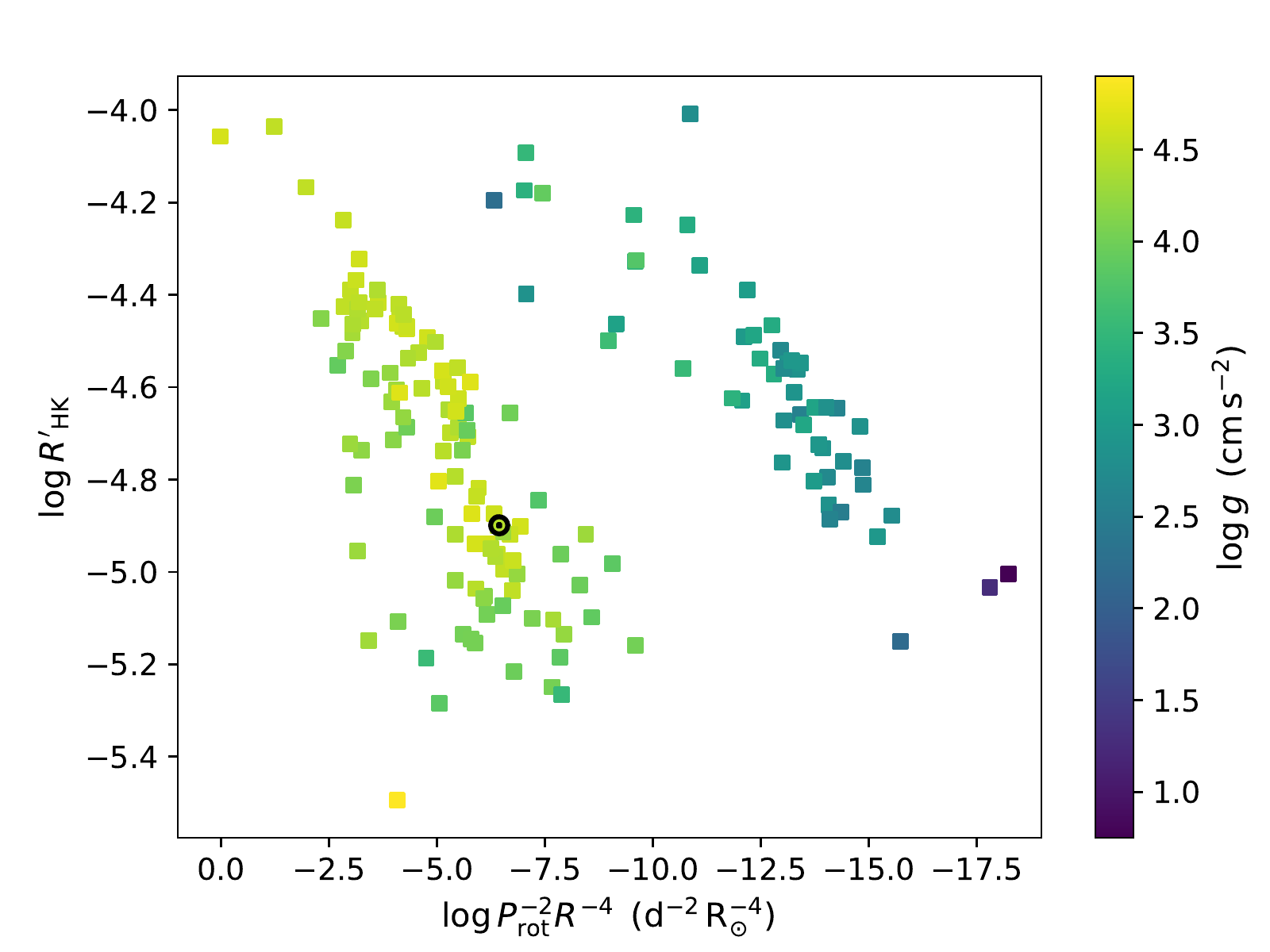}
\put(-355,235){\textbf{b}}
\caption{\textbf{Suggested alternative rotation--activity relations independent of convective turnover time.} \textbf{a}, Chromospheric Ca II H\&K luminosity, $\log L_{\rm HK}$, vs.~rotation period, $P_{\rm rot}$. \textbf{b}, Chromospheric activity, $\log R'_{\rm HK}$, vs.~combined rotation period and stellar radius, $P_{\rm rot}^{-2}R^{-4}$. The colour scale for $\log g$ is the same as in Figure \ref{rotact-fig}.}
\label{comparison-fig}
\end{figure}

In contrast, alternative activity scaling relations, $L_{\rm HK}$ vs.~$P_{\rm rot}$ and $R'_{\rm HK}$ vs.~$P_{\rm rot}^{-2}R^{-4}$, that remove the $\tau_{\rm c}$ dependence from the scaling\cite{Pizzolato2003Activity,Reiners2014Activity}, fail at closing the gap between the main sequence and evolved stars (Figure \ref{comparison-fig}). Both of these choices show a reasonably good rotation–activity relation for the main sequence stars, but the evolved stars form again their own separate clusters that do not coincide with the main sequence stars (Extended Data Figure \ref{cluster-fig} c,d).

A further quantitative comparison is possible between the $\log R'_{\rm HK}$ vs.~$\log \rm Ro$ and the $\log R'_{\rm HK}$ vs. $\log P_{\rm rot}^{-2}R^{-4}$ clustering results since in both cases it is possible to express the scatter of the clusters along $\log R'_{\rm HK}$. Against $\rm Ro$ we find the root mean square scatter to be $\rm rms = 0.114$ dex in $\log R'_{\rm HK}$ for the main cluster. This compares with a scatter of rms = 0.209 dex for the main sequence stars and $\rm rms = 0.120$ dex for the evolved stars against $P_{\rm rot}^{-2}R^{-4}$. While the difference is minimal for the evolved stars, the main sequence stars show notably worse scatter when scaled against $P_{\rm rot}^{-2}R^{-4}$.

All these results constitute a strong argument that the Rossby number provides a necessary, as well as sufficient, parametrisation for the stellar activity level over a wide range of evolutionary stages. The resulting scaling relation from the clustering analysis is $R'_{\rm HK} \propto {\rm Ro}^{-0.97}$. Using other proposed parameterisations still allows the construction of empirical rotation--activity relations for the main sequence stars, but these are of poorer quality and break down once more evolved stars are introduced into the picture. The fact that a unified scaling can only be achieved when both stellar rotation and convection are taken into account suggests that the underlying dynamos operating in these stars follow the turbulent dynamo paradigm. Moreover, also fully convective M-dwarfs have been shown to follow a common rotation–activity relation against Ro together with the more massive partially convective Solar-type main sequence stars\cite{Wright2016FullyConvective,Newton2017MDwarf,Wright2018DwarfRotationActivity}. It is thus reasonable to assume that there exists a universal dynamo scaling for all late type stars and that they all share the same underlying dynamo mechanism irrespective of their mass or evolutionary stage and the resulting vastly different internal structures.

\pagebreak

\begin{methods}

\subsection{Chromospheric activity and astrophysical parameters.}

The initial sample selection consists of all the stars included in the Mount Wilson Observatory (MWO) Calcium HK Project\cite{Wilson1978MWOHK} that have time series observations spanning over five years and covering at least four complete observing seasons. This selection consists of 224 stars in total. Using sufficiently extended time series data like this ensures averaging over yearly activity variations and facilitates a more reliable rotation period search. The final sample for which it was possible to determine both $P_{\rm rot}$ and $\tau_{\rm c}$ consists of 58 main sequence and 92 evolved stars. We supplemented this sample by five moderately to very active main sequence stars which have a poorer coverage in the MWO data but have accurate $P_{\rm rot}$ available from photometric studies\cite{Lehtinen2016Activity}.

The Ca II H\&K emission ratios, $R'_{\rm HK} = L_{\rm HK}/L_{\rm bol} = F'_{\rm HK}/\sigma T_{\rm eff}$, were calculated from the averaged MWO $S$-index observations\cite{Egeland2017MWO} after removing sections of data with apparent calibration issues and outliers more than $4\sigma$ away from the sample mean\cite{Olspert2018Cycle}. The conversion is colour dependent and defined separately for the main sequence and evolved stars\cite{Rutten1984CaHK}. The choice of appropriate conversion law was made based on the absolute $V$-band magnitude $M_V$, so that stars with $M_V$ more than 1 magnitude above the main sequence\cite{Allen2000Allen} were treated as evolved stars and the rest as main sequence stars. For the cooler stars with $B-V \ge 0.8$ (or $T_{\rm eff} \le 5300$ K) this procedure neatly separates the main sequence from the evolved stars (see Figure \ref{hr-fig}). For the hotter stars there is no clear separation between the different evolutionary stages in the Hertzsprung-Russell diagram, but this causes no issues for calculating $\log R'_{\rm HK}$ since the two conversion laws overlap in this region\cite{Rutten1984CaHK}.

$V$-band magnitudes and $B-V$ colours were adopted from the Hipparcos photometry\cite{esa1997hipparcos} for all stars apart from the Sun\cite{Allen2000Allen}. Parallaxes were drawn from the Gaia Data Release 2\cite{Gaia2016Gaia,Gaia2018DR2,Luri2018Parallax}, where available, otherwise adopting Hipparcos parallaxes\cite{vanLeeuwen2007Hipparcos}. Interstellar extinction was assumed to be $a_V = 1.5 \rm mag/kpc$ for calculating the absolute magnitudes\cite{Vergely1997Extinction}, which is a workable assumption since the stellar sample is located largely in our galactic neighborhood.

We adopted literature values for the effective temperature, $T_{\rm eff}$, and luminosity, $L$,\cite{Allen2000Allen,Andrae2018LumTeff} and the metallicity, [Fe/H],\cite{Gaspar2016Metallicity} of the stars. For some stars no values of $T_{\rm eff}$ or $L$ were available and these had to be estimated from the photometry\cite{Flower1996LumTeff} and the stellar radii, $R$, estimated thereof\cite{Gray2005Photospheres}. The Ca II H\&K luminosities were calculated from the emission fluxes as $L_{\rm HK} = 4\pi R^2 F'_{\rm HK}$. Values of surface gravity, $\log g$, were compiled from the PASTEL catalogue\cite{Soubiran2016PASTEL}.

\subsection{Stellar structure models.}

Since the convective turnover time $\tau_{\rm c}$ depends on the stellar mass, chemical composition, and evolutionary stage, we derived its value for each star individually using a grid of stellar evolution models from the Yale-Potsdam Stellar Isochrones\cite{Spada_ea2017} (YaPSI). In the YaPSI models, $\tau_{\rm c}$ is calculated according to a global definition\cite{KimDemarque1996} based on an average over the whole convection zone of the star.

The position of each star in the Hertzsprung-Russell diagram (i.e., $\log T_{\rm eff}$, $\log L/L_\odot$) was compared with the model evolutionary tracks, and a subset of tracks matching its parameters within the observational uncertainties was selected. We assumed uncertainties of 100 K in $T_{\rm eff}$ and 0.12 dex in $\log L/L_\odot$. A best-fitting track was then generated from the selected tracks by linear interpolation, and $\tau_{\rm c}$ was extracted from this synthetic track at the point of closest approach to the observed parameters. We repeated this procedure for three different metallicities: [Fe/H] = −0.5, 0.0, and 0.3 (with initial helium fraction kept constant at Y = 0.28), obtaining an estimate of $\tau_{\rm c}$ in each case. This range of [Fe/H] encompasses almost the totality of the stars within our sample (Extended Data Figure \ref{feh-fig}). The final value of $\tau_{\rm c}$ was obtained by linear interpolation in [Fe/H]. The range of $\tau_{\rm c}$ obtained from the models at different metallicities was used to estimate the uncertainty on $\tau_{\rm c}$. It should be noted that although this uncertainty is only qualitative (i.e., it cannot be interpreted as a formal error bar on $\tau_{\rm c}$), we found that the stars with the largest uncertainty are those close to the $\tau_{\rm c} \approx 0$ d limit, lending further support to excluding the stars with the smallest $\tau_{\rm c}$ from the analysis. The uncertainties of $\rm Ro$, as shown in Figure \ref{evo-fig}, were computed by propagation of uncertainty from the $\tau_{\rm c}$ uncertainties and the standard deviations of the $P_{\rm rot}$ estimates.

For stars on the main sequence, our theoretical values of $\tau_{\rm c}$ are in good qualitative agreement with the classical empirical estimates\cite{Noyes_ea1984}, except for a roughly $\tau_{\rm c}/\tau_{\rm c,empirical} = 2.6$ factor between the two. For stars in the subgiant and red giant  branch phases, the stellar evolution models predict a strong dependence of $\tau_{\rm c}$ on the evolutionary stage of the star, which is not captured by the empirical estimates.

The evolutionary stages of the stars, indicated in Figure \ref{evo-fig}, were determined from the evolutionary track fits using the following criteria. The termination age main sequence occurs when the hydrogen mass fraction in the core of a star reaches below $10^{-4}$, after which the stars enter their subgiant phase. The transition from the subgiant to the giant phase was set to occur at the bottom of the red giant branch, which was defined to be reached once the inert helium core reaches a mass of $M_{\rm He} > 0.1M_\odot$.

Determining precise stellar properties for red giants by matching their observed properties to evolutionary tracks and isochrones is notoriously difficult\cite{Serenelli:2013,SilvaAguirre:2016}. In addition, a significant source of scatter in the value of $\tau_{\rm c}$ for subgiant and early red giant stars comes from their fast evolutionary timescales, and is therefore unfortunately unavoidable. This effect is illustrated in Extended Data Figure \ref{tauc_evol}, which shows a selection of evolution tracks for the main sequence and post-main sequence phases and the corresponding evolution of $\tau_{\rm c}$. For stars of mass $M < 1.3 M_\odot$, $\tau_{\rm c}$ remains essentially constant during the main sequence and increases smoothly during the post-main sequence evolution. On the other hand, stars of mass $M > 1.5 M\odot$ have no outer convection zone in the main sequence. As they reach the bottom of the red giant branch, a sizable convective envelope develops in a much faster timescale than in their less massive counterparts. Correspondingly, $\tau_{\rm c}$ changes very rapidly from zero to a few hundred days, a value typical of red giant stars.

\subsection{Period analysis.}

The $P_{\rm rot}$ were determined for the stars from the rotational modulation in the MWO $S$-index time series, caused by the transit of chromospheric active regions over the visible stellar hemispheres\cite{Olspert2018Cycle}. An initial period search was performed for each complete observing season using periodic Gaussian processes\cite{Wang2012GP}. If the different seasons yielded repeatedly comparable period values, these were used as initial guesses for computing the final period estimate from the full time series using the Continuous Period Search method\cite{Lehtinen2011CPS}.

\subsection{Gaussian clustering.}

The clustering analysis of the rotation--activity diagrams was done using the Gaussian Mixture Model with Expectation Maximisation\cite{BarberBRML2012}. We searched for the statistically most likely configuration of clusters, not assuming any prior knowledge neither about the number of clusters nor their covariances. The algorithm was run for the number of clusters ranging from one to five and the optimal configuration was determined by minimising the Bayesian Information Criterion\cite{Stoica2004ModelOrder}. The most probable cluster membership was determined afterwards for the individual stars using Mahalanobis distance\cite{Mahalanobis1936distance}.

\subsection{Data availability.}

The Mount Wilson Observatory HK Project data are available online at \\ \texttt{ftp://solis.nso.edu/MountWilson\_HK/} and the Gaia Data Release 2 from the Gaia Archive at \texttt{http://gea.esac.esa.int/archive/}. The YaPSI stellar models are available at \\ \texttt{http://www.astro.yale.edu/yapsi/}. The adopted and derived astrophysical parameters for the stellar sample used in the present study are available in online tables at the CDS astronomical data center via anonymous ftp to \texttt{cdsarc.u-strasbg.fr} (130.79.128.5) or via \\ \texttt{http://cdsarc.u-strasbg.fr/viz-bin/qcat?J/other/NatAs}.

\end{methods}

\bibliographystyle{naturemag}
\bibliography{rotact}

\pagebreak

\begin{addendum}
\item This work has made use of data from the European Space Agency (ESA) mission {\it Gaia} (\texttt{https://www.cosmos.esa.int/gaia}), processed by the {\it Gaia} Data Processing and Analysis Consortium (DPAC, \texttt{https://www.cosmos.esa.int/web/gaia/dpac/consortium}). Funding for the DPAC has been provided by national institutions, in particular the institutions participating in the {\it Gaia} Multilateral Agreement.

The chromospheric activity data derive from the Mount Wilson Observatory HK Project, which was supported by both public and private funds through the Carnegie Observatories, the Mount Wilson Institute, and the Harvard-Smithsonian Center for Astrophysics starting in 1966 and continuing for over 36 years.  These data are the result of the dedicated work of O. Wilson, A. Vaughan, G. Preston, D. Duncan, S. Baliunas, and many others.

The work has made use of the SIMBAD database at CDS, Strasbourg, France, and NASA's Astrophysics Data System (ADS) services.

J.J.L. acknowledges financial support from the Independent Max Planck Research Group ``SOLSTAR''. F.S. acknowledges the support of the German space agency (Deutsches Zentrum f\"ur Luft- und Raumfahrt) under PLATO Data Center grant 50OO1501. M.J.K., N.O., and P.J.K. acknowledge the support of the Academy of Finland ReSoLVE Centre of Excellence (grant number 307411). P.J.K. acknowledges support from DFG Heisenberg grant (No. KA 4825/2-1). This project has received funding from the European Research Council (ERC) under the European Union's Horizon 2020 research and innovation programme (Project UniSDyn, grant agreement n:o 818665).

\textbf{Author contributions} All authors contributed to the research and its design. J.J.L. and N.O. led the data analysis of the observations. F.S. led the stellar structure modelling. M.J.K. and P.J.K. led the theoretical interpretation of the obtained results. All authors contributed to the discussion of the results and to the manuscript.

\textbf{Competing interests} The authors declare no competing interests.

\textbf{Author information} Reprints and permissions information is available at \texttt{www.nature.com/reprints}. Correspondence and requests for materials should be addressed to J.J.L. (lehtinen@mps.mpg.de).
\end{addendum}

\renewcommand{\figurename}{Extended Data Figure}
\setcounter{figure}{0}

\begin{figure}
\centering
\begin{tabular}{cc}
\includegraphics[width=.5\linewidth]{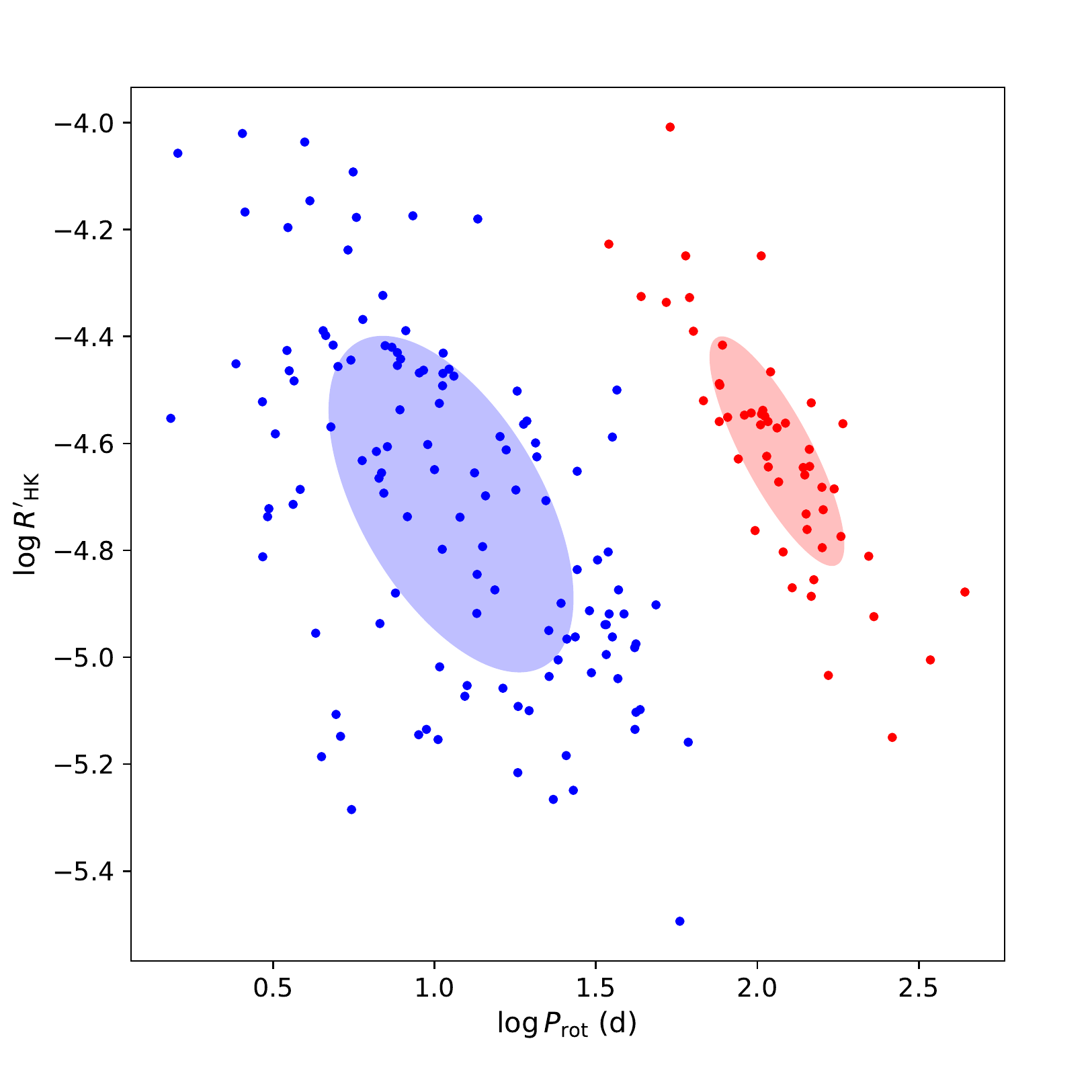}
\put(-240,205){\textbf{a}} &
\includegraphics[width=.5\linewidth]{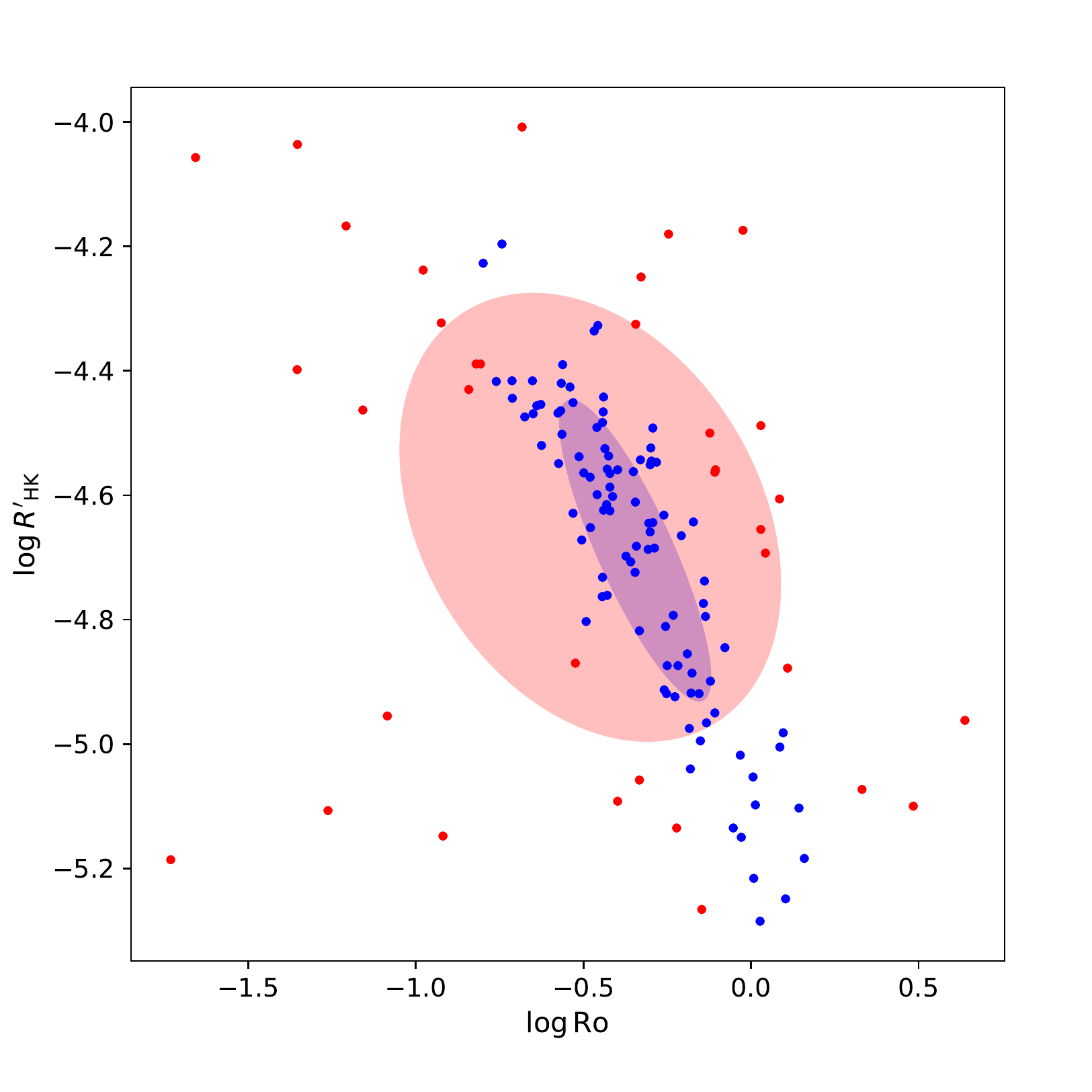}
\put(-240,205){\textbf{b}} \\
\includegraphics[width=.5\linewidth]{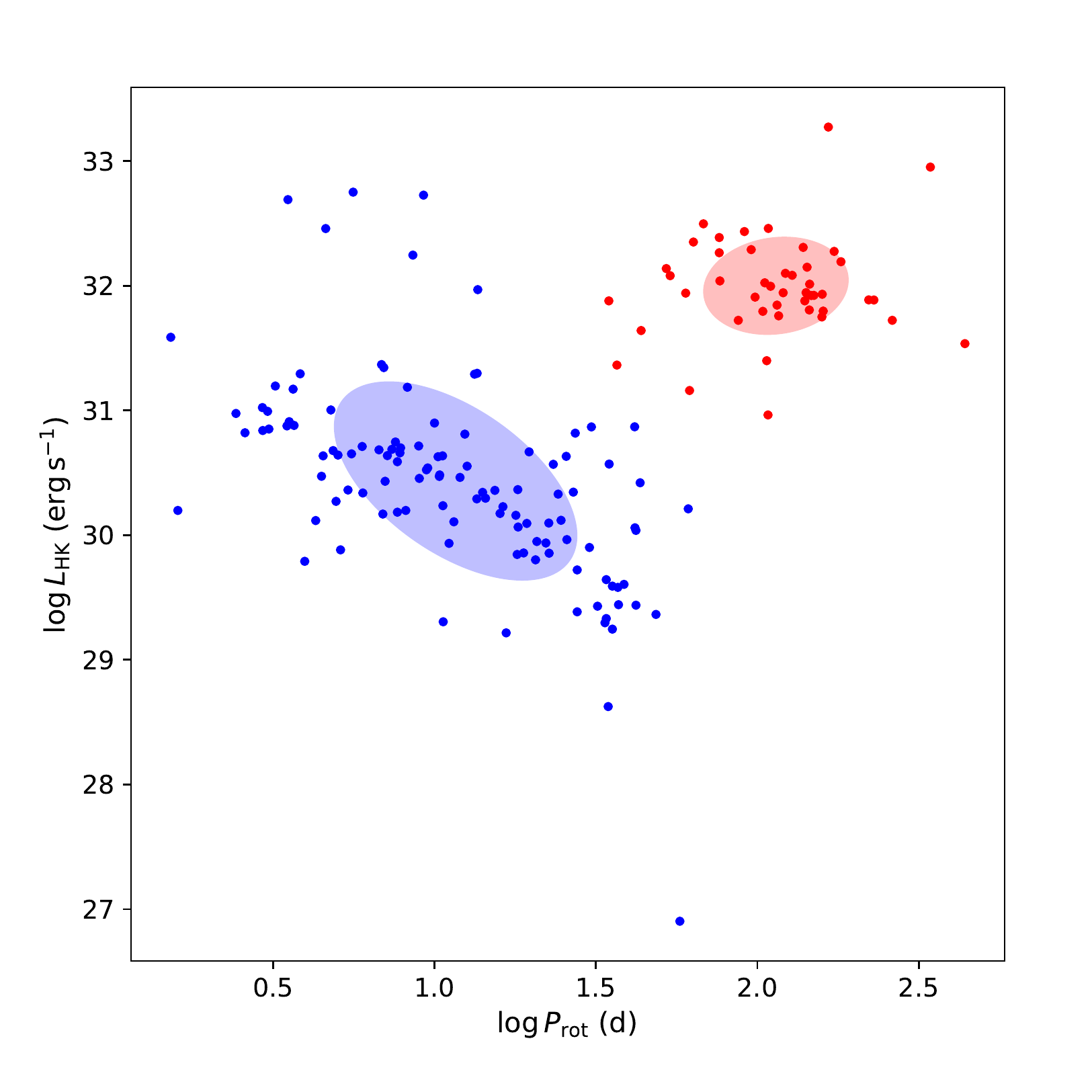}
\put(-240,205){\textbf{c}} &
\includegraphics[width=.5\linewidth]{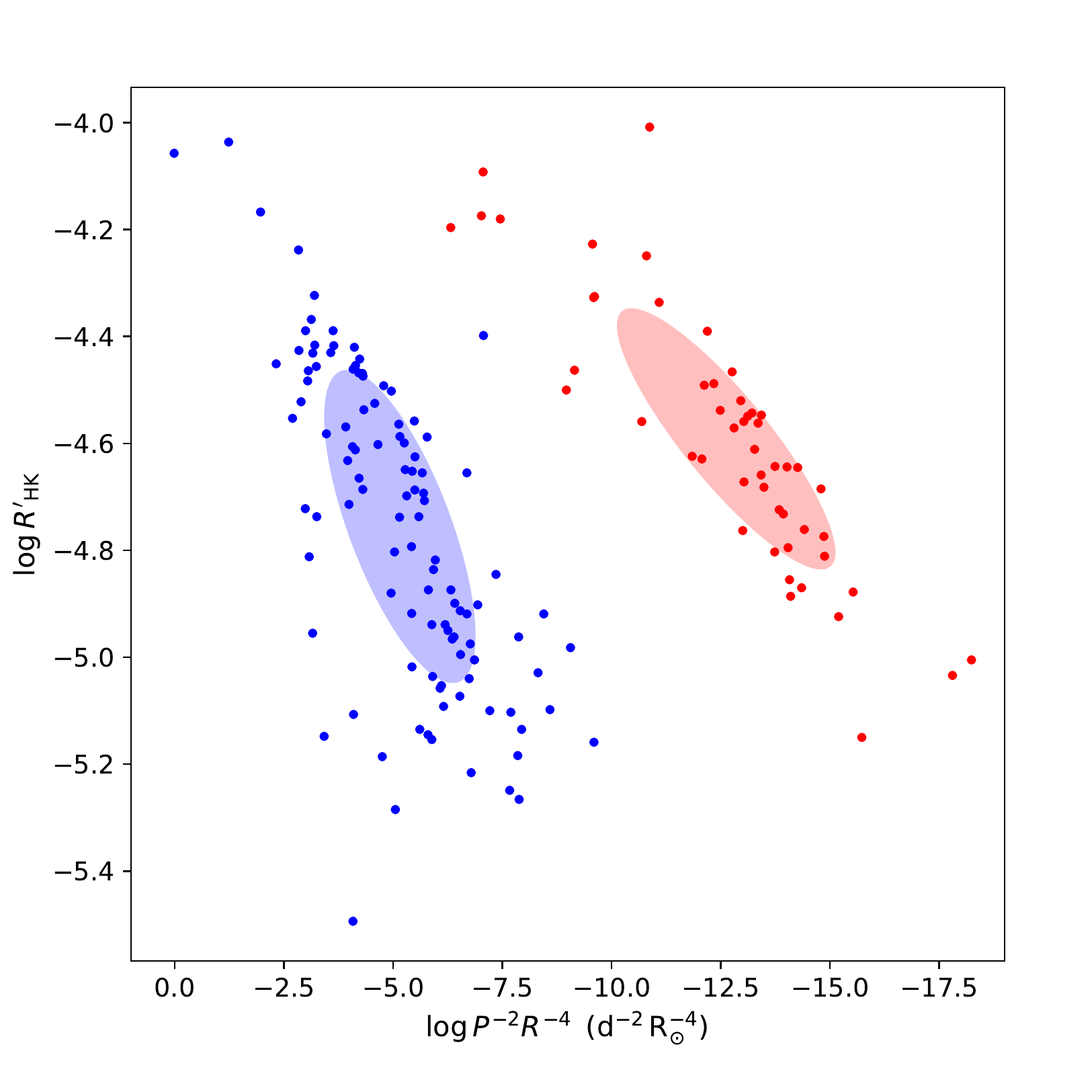}
\put(-240,205){\textbf{d}}
%\multicolumn{2}{c}{
%\includegraphics[width=.5\linewidth]{rotact-pizzolato_fig.pdf}
%\put(-240,205){\textbf{c}}
%}
\end{tabular}
\caption{\textbf{Gaussian clustering for the rotation--activity relation.} \textbf{a}, Chromospheric activity, $\log R'_{\rm HK}$, vs.~rotation period, $\log P_{\rm rot}$. \textbf{b}, Chromospheric activity, $\log R'_{\rm HK}$, vs.~Rossby number, $\log \rm Ro$. \textbf{c}, Chromospheric Ca II H\&K luminosity $\log L_{\rm HK}$ vs.~rotation period, $\log P_{\rm rot}$. \textbf{d}, Chromospheric activity, $\log R'_{\rm HK}$, vs.~combined rotation period and stellar radius, $\log P_{\rm rot}^{-2}R^{-4}$. Optimal clustering of the data is indicated by the blue and red ellipses, reflecting the corresponding 95\% confidence regions. The individual stars are coloured according to their inferred cluster membership.}
\label{cluster-fig}
\end{figure}

\begin{figure}
\centering
\includegraphics[width=\linewidth]{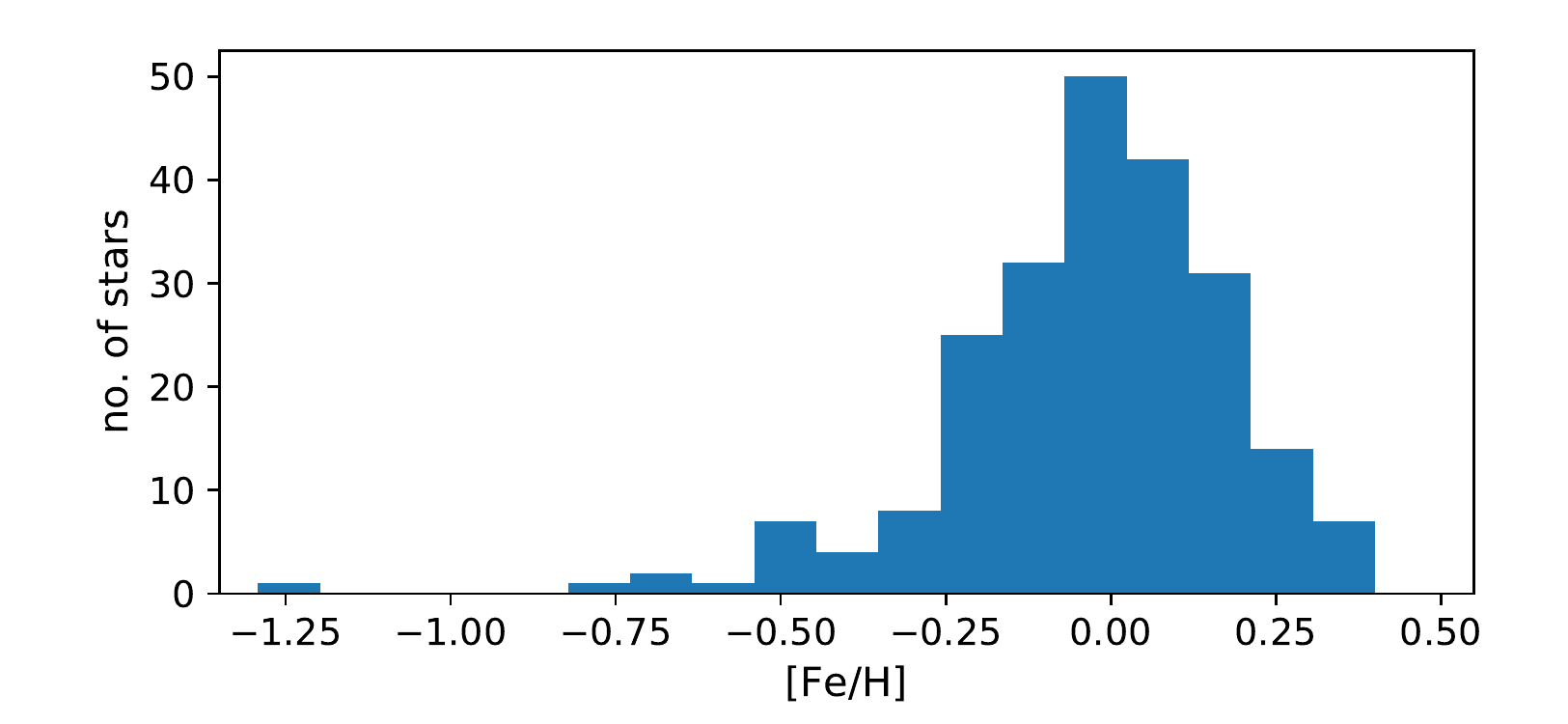}
\caption{\textbf{Metallicity distribution of the sample stars.}}
\label{feh-fig}
\end{figure}

\begin{figure}
\centering
\includegraphics[width=\linewidth]{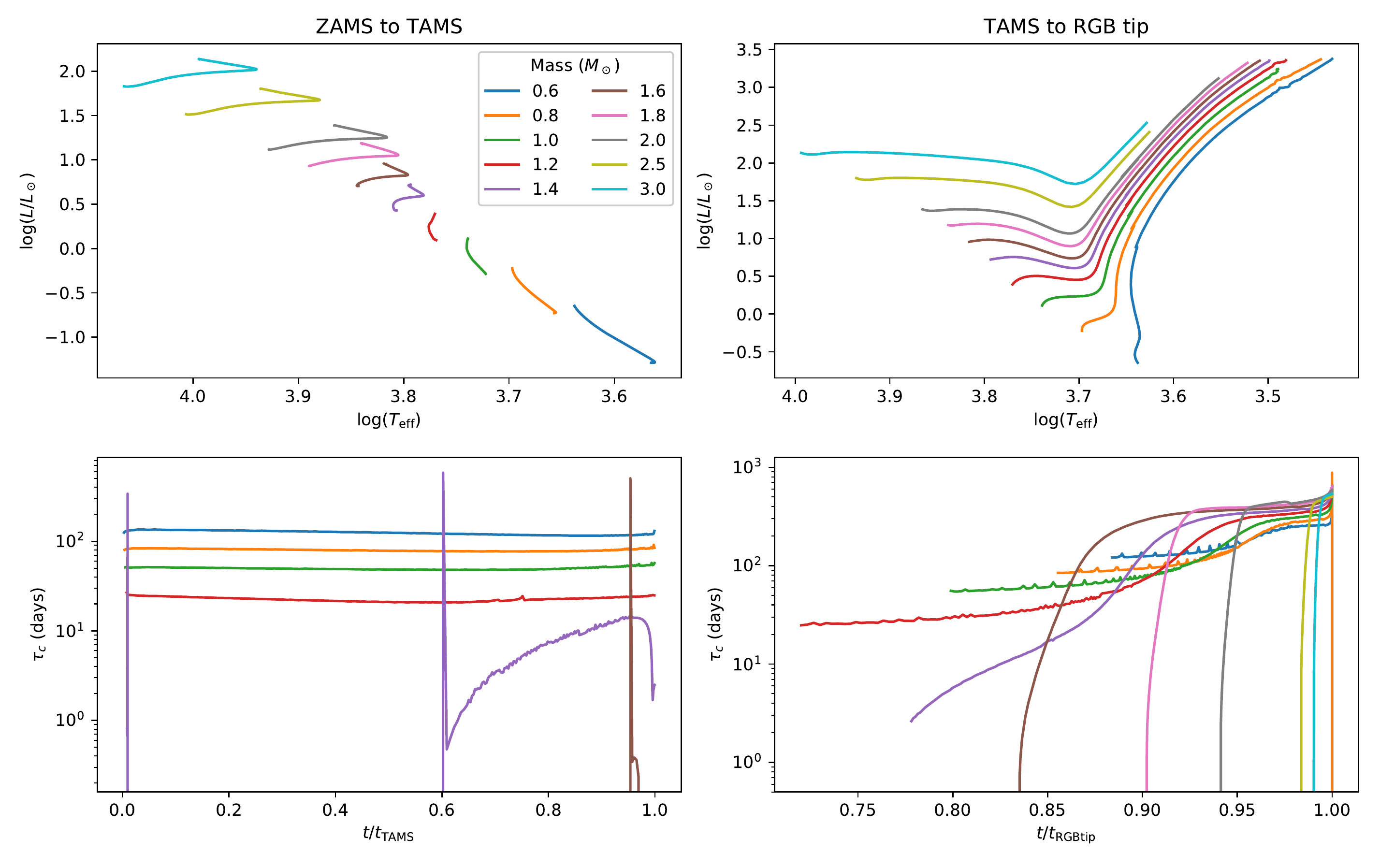}
\put(-462,275){\textbf{a}}
\put(-230,275){\textbf{b}}
\put(-462,135){\textbf{c}}
\put(-230,135){\textbf{d}}
\caption{\textbf{Selected stellar evolution tracks with the time evolution of convective turnover time, $\tau_{\rm c}$.} \textbf{a}, Evolution in the Hertzsprung-Russell (HR) diagram from the zero age main sequence (ZAMS) to the termination age main sequence (TAMS). \textbf{b}, Evolution in the HR diagram from TAMS to the red giant branch (RGB) tip. \textbf{c}, Evolution of $\tau_{\rm c}$ from ZAMS to TAMS. \textbf{d}, Evolution of $\tau_{\rm c}$ from TAMS to RGB tip. All ages are normalised to TAMS or RGB tip age.}
\label{tauc_evol}
\end{figure}

\end{document}